To appear in *Nature* on 21st March 2010

# Intense Star Formation within Resolved Compact Regions in a Galaxy at z=2.3

A. M. Swinbank[1], I. Smail[1], S. Longmore[2], A. I. Harris[3], A. J. Baker[4], C. De Breuck[5], J. Richard[1], A. C. Edge[1], R. J. Ivison[6,7], R. Blundell[2], K. Coppin[1], P. Cox[8], M. Gurwell[2], L. J. Hainline[3], M. Krips[8], A. Lundgren[9], R. Neri[8], B. Siana[10], G. Siringo[9], D. P. Stark[11], D. Wilner[2], J. D. Younger[12,**]

[1] *Institute for Computational Cosmology, Durham University, South Road, Durham DH1 3LE, UK; Email: a.m.swinbank@dur.ac.uk*
[2] *Harvard-Smithsonian Center For Astrophysics, 60 Garden Street, Cambridge, MA, 02138, USA*
[3] *Department of Astronomy, University of Maryland, College Park, MD 20742, USA*
[4] *Department of Physics and Astronomy, Rutgers, University of New Jersey, 136 Frelinghuysen Road, Piscataway, NJ 08854-8019, USA*
[5] *European Southern Observatory, Karl-Schwarzschild Strasse, 85748 Garching bei München, Germany*
[6] *UK Astronomy Technology Centre, Science and Technology Facilities Council, Royal Observatory, Blackford Hill, Edinburgh EH9 3HJ*
[7] *Institute for Astronomy, Royal Observatory of Edinburgh, Blackford Hill, Edinburgh EH9 3HJ*
[8] *Institut de Radio Astronomie Millimétrique, 300 Rue de la Piscine, Domaine Universitaire, 38406 Saint Martin d'Héres, France*
[9] *European Southern Observatory, Alonso de Cordova 3107, Cassilla 19001, Santiago 19, Chile*
[10] *California Institute of Technology, MS 105-24, Pasadena, CA 91125, USA*
[11] *Institute of Astronomy, University of Cambridge, Madingley Road, Cambridge CB3 0HA, USA*
[12] *Institute for Advanced Study, Einstein Drive, Princeton NJ 08540, USA ** HubbleFellow*

**Massive galaxies in the early Universe have been shown to be forming stars at surprisingly high rates[1—3]. Prominent examples are dust-obscured galaxies which are luminous when observed at sub-millimeter (sub-mm) wavelengths and which may be forming stars at a rate of 1,000 solar masses ($M_\odot$) per year[4—6]. These intense bursts of star formation are believed to be driven by mergers between gas rich galaxies[7—9]. However, probing the properties of individual star-forming regions within these galaxies is beyond the spatial resolution and sensitivity of even the largest telescopes at present. Here, we report observations of the sub-mm galaxy SMMJ2135-0102 at redshift *z*=2.3259 which has been gravitationally magnified by a factor of 32 by a massive foreground galaxy cluster lens. This cosmic magnification, when combined with high-resolution sub-mm imaging, resolves the star-forming regions at a linear scale of just ~100 parsecs. We find that the luminosity densities of these star-forming regions are comparable to the dense cores of giant molecular clouds in the local Universe, but they are ~100× larger and $10^7$ times more luminous. Although vigorously star-forming, the underlying physics of the star formation processes at *z*~2 appears to be similar to that seen in local galaxies even though the energetics are unlike anything found in the present-day Universe.**

Strong gravitational lensing, light bent by massive galaxy clusters, magnifies the images of distant galaxies which serendipitously lie behind them, offering a direct route to probing the physical processes occurring within star-forming regions in high-redshift galaxies. During an 870μm APEX/LABOCA observation of the massive galaxy cluster MACSJ2135-010217 ($z_{cl}$=0.325), we recently discovered a uniquely bright galaxy with an 870μm flux of 106.0±7.0mJy (Fig. 1). The optical counterpart is faint, with $I_{AB}$=23.6±0.2 and $K_{AB}$=19.77±0.07, and is extended along a roughly East-West direction, consistent with it being a gravitationally lensed background galaxy. The mid- and far-infrared colours ($S_{24}/S_{70}$=0.4±0.2) and red optical/near-infrared colours also suggest the galaxy lies beyond the cluster at *z*>1.5[10] (Supplementary Information) and indeed carbon monoxide (CO) *J*=1-0 emission at 34.64GHz unambiguously identified the redshift as *z*=2.3259±0.0001 (Fig. 2). With source and lens redshifts known, we used the gravitational lens model of the galaxy cluster (Supplementary Information) to correct for the lensing distortion, deriving an amplification factor for the background galaxy of $\mu$=32.5±4.5.

Observations of molecular and continuum emission provide gas and stellar mass estimates. The observed velocity integrated flux in CO(1-0) is $f_{CO}$=2.3±0.1Jy km/s, the CO(3-2)/CO(1-0) flux ratio of 5.9±0.3 suggests that the molecular gas is subthermally excited (Fig 2). Assuming a cosmology with $\Omega_\Lambda$=0.73, $\Omega_m$=0.27, and $H_0$=72km/s/Mpc$^{-1}$, and a CO-$H_2$ conversion factor $\alpha$=0.8 (K.km/s/pc$^2$)$^{-1}$ (which is appropriate for the smoothly distributed, high-pressure, largely molecular inter-stellar medium with subthermal CO excitation[9,13,14]) we derive a cold gas mass of $M_{gas}=M(H_2+He)=\alpha L'_{CO(1-0)}$=1.6±0.1x10$^{10}$ solar masses ($M_\odot$). We estimate the stellar mass by fitting stellar population synthesis models to the rest-frame UV—near-infrared spectral energy distribution (SED)[15] shown in Fig. 3. The best-fit SEDs have a range of ages from 10—30Myr with significant dust extinction,

$E(B-V)$=1.0±0.1, and a stellar mass (corrected for lensing) of $M_{stars}$=3±1x10$^{10}$$M_{\odot}$. Taken together, these imply a baryonic mass of $M_{bary}=M_{gas}+M_{stars}$= 4±2×10$^{10}$$M_{\odot}$, with approximately 35% of this in cold molecular gas.

Rest-frame far-infrared radiation from dust-reprocessed UV light provides an extinction-free measure of the instantaneous star formation rate of a galaxy. Correcting for lens magnification, the intrinsic observed-frame 870μm flux is $S_{870\mu m}$=3.0±0.4mJy, suggestive of a typical high-redshift Ultra-Luminous Infrared Galaxy (ULIRG)[1,3,12]. Observations at 350μm with APEX/SABOCA and 434μm with the Sub-Millimeter Array (SMA) constrain the far-infrared SED (Fig. 3). Using a modified blackbody spectrum[3] with a two-component dust model (with $T_d$=30K and 60K) we derive a bolometric luminosity (corrected for lensing amplification) of $L_{bol}$=1.2±0.2×10$^{12}$ solar luminosities ($L_{\odot}$), suggesting a star-formation rate of $SFR$=210±50$M_{\odot}/yr$[16]. If this star formation rate has been continuous, it would take just ~150Myr to build the current stellar mass; the remaining gas depletion timescale would be a further 75Myr, suggesting that the intense star formation episode we observe may be the first major growth phase of this galaxy. To set the global properties of the galaxy in the context of other galaxy populations, it is also possible to calculate the efficiency at which the dense gas is converted into stars. The theoretical limit at which stars can form[17] is given by $SFR=\varepsilon M_{gas}/t_{dyn}$ where $\varepsilon$ is the star- formation efficiency, and $t_{dyn}$ is the dynamical (free—fall) time which is given by $t_{dyn}=(r^3/2GM)^{0.5}$. Adopting $r$=1.5kpc the star formation efficiency is $\varepsilon$~0.02, which is consistent with that derived for local ULIRGs[18] and archetypal high-redshift SMGs[9], but a factor 20 lower than the most extreme 'hyper'-starbursts at z~6[19].

SMA observations spatially resolve the galaxy's 870μm (345GHz) continuum emission with a 0.2" synthesized beam, providing a detailed view of the galaxy's morphology. Fig. 1 shows eight discrete components over ~4" in projection. These represent two mirror images of the source, each comprising four separate emission regions, reflected about the lensing critical curve. The map contains a total flux of $S_{850\mu m}$=86±3mJy, or 82±2% of the flux in the LABOCA map, suggesting that the structures in the SMA map contain the bulk of the 870μm luminosity. Reconstructing the source-plane image, the galaxy comprises four bright star-forming regions in the source plane (A,B,C,D), which are separated by 1.5kpc in projection (A&B separated by ~800pc; C&D by ~450pc). Assuming the dynamics of the CO emission trace the virialized potential well of the galaxy, then on these scales the dynamical mass of the system is $M_{dyn}$~4—8x10$^{10}$$M_{\odot}$, in good agreement with the gas and stellar mass estimates.

For the most highly amplified components (D1/D2, Fig. 1), the source-plane resolution reaches ~90 parsecs, only slightly larger than the ~60 parsec characteristic size of giant molecular clouds (GMCs) in the Milky Way[20]. This is consistent with the black-body radius ($R_{bb}$) estimated from the bolometric luminosity and dust temperature ($T_d$) via the scaling relation $L_{bol}/L_{\odot}=(R_{bb}/R_{\odot})^2(T_d/T_{\odot})^4$, where $R_{bb}$ is the physical black-body radius and $T_{\odot}$ denotes the solar temperature)[21]. Taking $L_{bol}$=0.6—1.1x10$^{12}$ $L_{\odot}$, and assuming characteristic dust temperatures of $T_d$=30—60K for each of the star-forming regions within SMMJ2135-0102, the predicted sizes are $R_{bb}$~100—300pc. This is consistent with those measured on the sky at 870μm.

Given that the star forming regions in SMMJ2135-0102 are similar in size to GMCs in the Milky-Way and local group galaxies, it is instructive to see how the luminosities compare. Within typical star-forming regions, constant energy density produces a correlation between size and luminosity such that $L_{260}\propto r^3$ which appears to hold over several orders of magnitude, from 1—100pc[20-24]. Within the dense central cores of actively star-forming GMCs, however, luminosity from massive stars dominate across the 1pc regions in which they have formed, producing luminosity densities a factor ~100× higher than averaged across GMCs[25]. We therefore plot two lines with slope 3, meaning a constant energy density, with the lower one roughly centered on the GMCs, and the upper one a factor of 100 higher. The star-forming regions within SMMJ2135-0102 are ~100pc across, two orders of magnitude larger than the 1pc average for dense GMC cores, but as Fig. 4 shows, their luminosities are approximately 100x higher than expected for typical star-forming regions of comparable size in the low-redshift Universe. This likely means that the variable is the number of star forming cores, so a region in SMMJ2135-0102 that has a size of ~100 pc contains ~10$^7$ 1-pc-sized cores[25—27]. The luminosity- (and therefore star formation-) density of the star-forming regions within SMMJ2135-0102 are also similar to those found in the highest density regions of the local starburst galaxy Arp220, although they are scaled up by a factor 10 in both size and luminosity[28] (Supplementary Information). Thus, while the energetics of the star-forming regions within SMMJ2135-0102 are unlike anything found in the present-day Universe, the relation between size and luminosity is similar to local, dense GMC cores, suggesting that the underlying physics of the star-forming processes is similar. These results suggest that the recipes developed to understand star-forming processes in the Milky Way and local galaxies can be used to model the star formation processes in these high-redshift galaxies.

**Acknowledgements**. We gratefully acknowledge the staff at Green Bank Telescope for scheduling the Zpectrometer observations at short notice and the ESO director for granting DDT observations with SABOCA. AMS gratefully acknowledges a Royal Astronomical Society Sir Norman Lockyer Fellowship, and JR and DPS acknowledge a Marie Curie fellowship and STFC fellowship respectively. JDY acknowledges support from NASA through a Hubble Fellowship. The Zpectrometer observations were carried out under program 09A-040 on the Robert C. Byrd Green Bank Telescope, which is operated by the National Radio Astronomy Observatory, a facility of the National Science Foundation operated under cooperative agreement by Associated Universities, Inc. The APEX and FORS observations were carried out with ESO Telescopes under programs 083.B-0941, 283.A-5032 and 078.A-0420. APEX is a collaboration between the Max Planck Institute for Radio Astronomy, the Onsala Space Observatory and ESO. The SMA is a joint project between the Smithsonian Astrophysical Observatory and the Academia Sinica Institute of Astronomy and Astrophysics and is funded by the Smithsonian Institution and the Academia Sinica. The CO(3-2) observations were carried out with the IRAM PdBI which is supported by INSU/CNRS (France), MPG (Germany), and IGN (Spain).

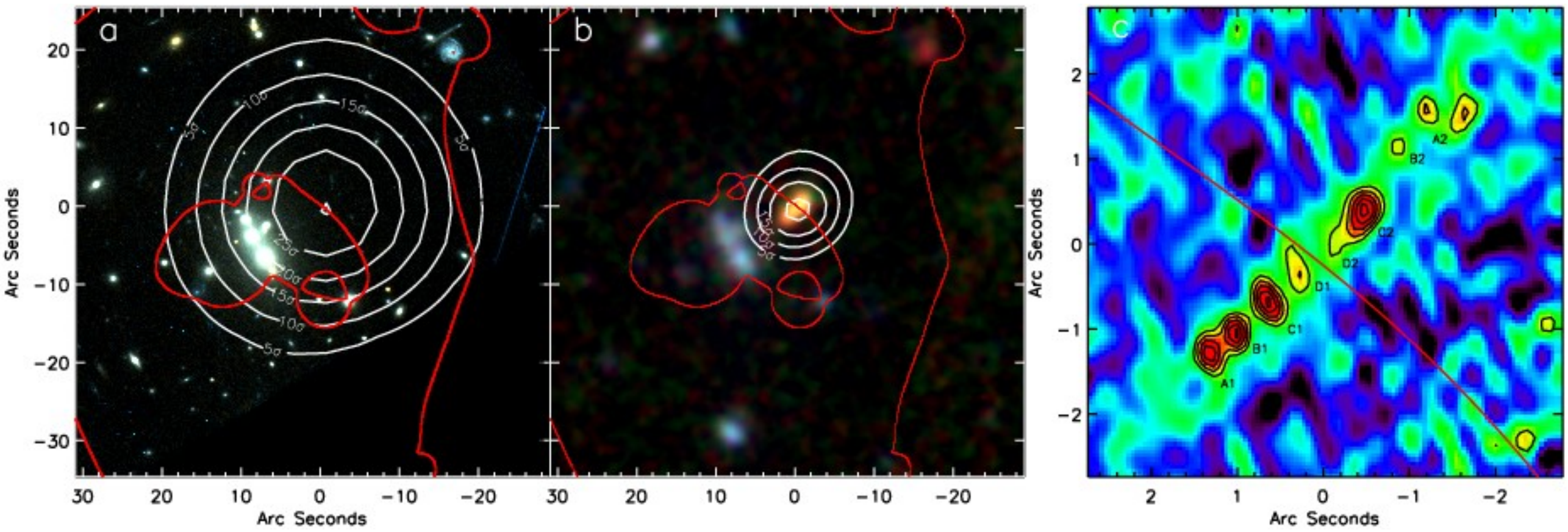


**Figure 1. Multi-wavelength images of the galaxy cluster. a**, Hubble Space Telescope VI-band colour image of the galaxy cluster with white contours denoting the 870μm emission from observations with Large Apex Bolometer Camera (LABOCA) on the Atacama Pathfinder Experiment (APEX) telescope. Contours denote 5, 10, 15, 20, 25 & 30σ (r.m.s. noise: 3.5mJy) identifying a sub-mm galaxy with flux 106.0±7.0mJy (the quoted error on the galaxy flux includes calibration uncertainties) at $\alpha$:21:35:11.6 $\delta$:-01:02:52.0 (J2000). The optical counter part is faint with $I_{AB}$=23.6±0.2. The solid red lines denote the $z$=2.326 radial and tangential critical curves from the best-fit lens model. **b,** True colour IRAC 3.6, 4.5, 8.0μm image of the cluster core with contours denoting the 350μm emission from APEX/SABOCA. Contours are spaced at 5, 10, 15, & 20σ (r.m.s. noise: 23mJy); the 350μm flux is 530±60mJy. The mid-infrared counterpart is clearly visible as an extended red galaxy centered at the sub-mm position. The LABOCA and SABOCA FWHM beams are 19" and 8" respectively. The origins of both images are on the lensed galaxy with North up and East left. **c,** SMA 870mm image of the galaxy. The map shows eight individual components, separated by up to 4" in projection. The red line is the same $z$=2.326 radial critical curve as in panels the top two panels. Components (A,B,C,D) represent two mirror images of the galaxy, each comprising four separate emission regions reflected about the lensing critical curve. The lower-right inset shows the 0.33"x0.21" synthesised beam with position angle of 15° East of North. The contours start at 3σ and are spaced by 1σ.

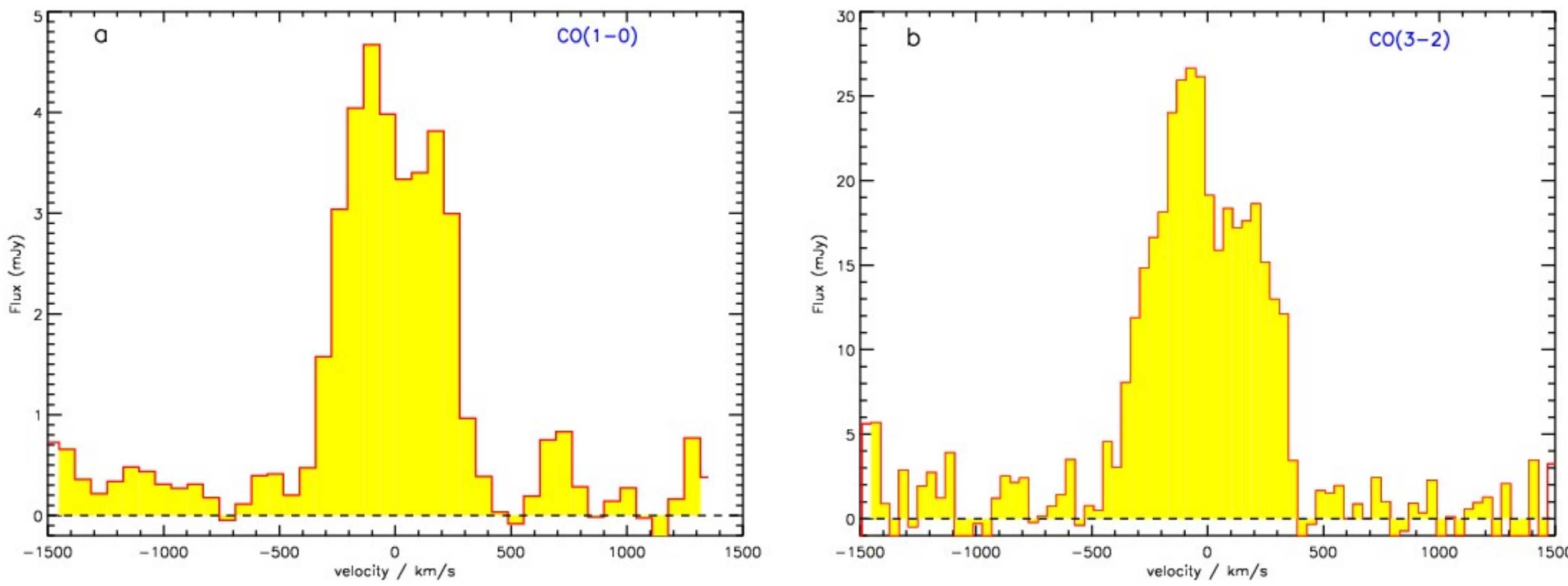


F**igure 2. Carbon monoxide (CO) observations of SMMJ2135-0102 obtained with the Zpectrometer on Green Bank Telescope (top) and Plateau de Bure Interferometer.** The redshift of is $z$=2.3259±0.0001 was derived from observations using Zpectrometer, a wide-band spectral instrument on the Green Bank Telescope[11]. **a**, Zpectrometer CO(1-0) spectrum, showing a double-horned profile with a velocity offset of 290±30km/s between the two peaks. **b**, Plateau de Bure observations of the CO(3-2) emission, confirming both the redshift and the multiple velocity components seen in CO(1-0).

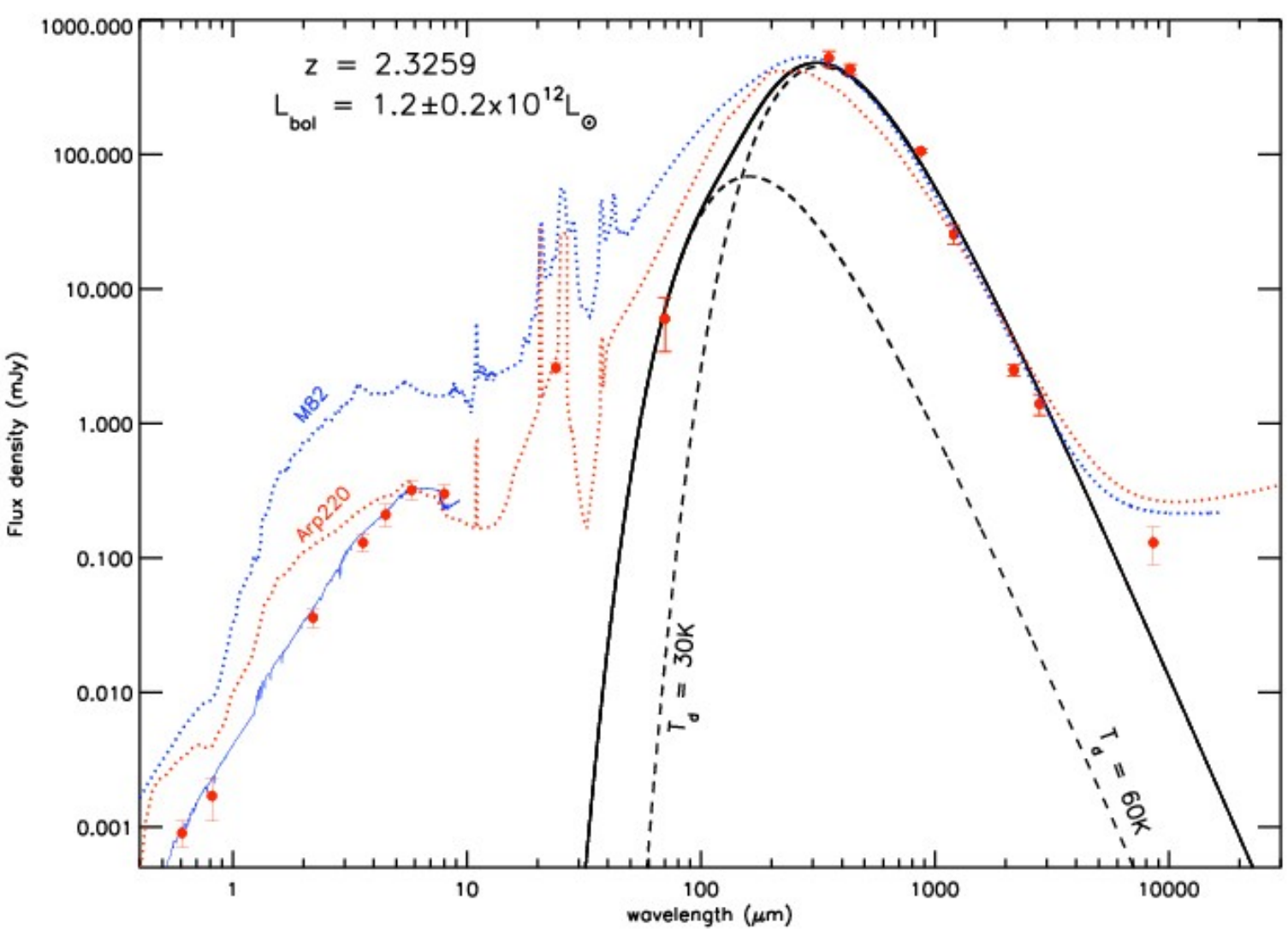


**Figure 3. Spectral energy distribution (SED) of the lensed galaxy.** To illustrate how the SED of the galaxy compares to local starbursts, we overlay the SEDs from M82 (blue dashed line) and Arp220 (red dashed line), both scaled to equal flux at 1.2mm. The solid black line denotes the best fit spectrum, a two component dust model with temperatures fixed at $T_d$=30 and 60K and a dust emissivity with $\beta$=1.8 in a modified black body function. We also overlaid the best-fit stellar SED to the optical to mid-infrared photometry (solid blue line) from which we estimate stellar age, extinction, luminosity, and mass through population synthesis.

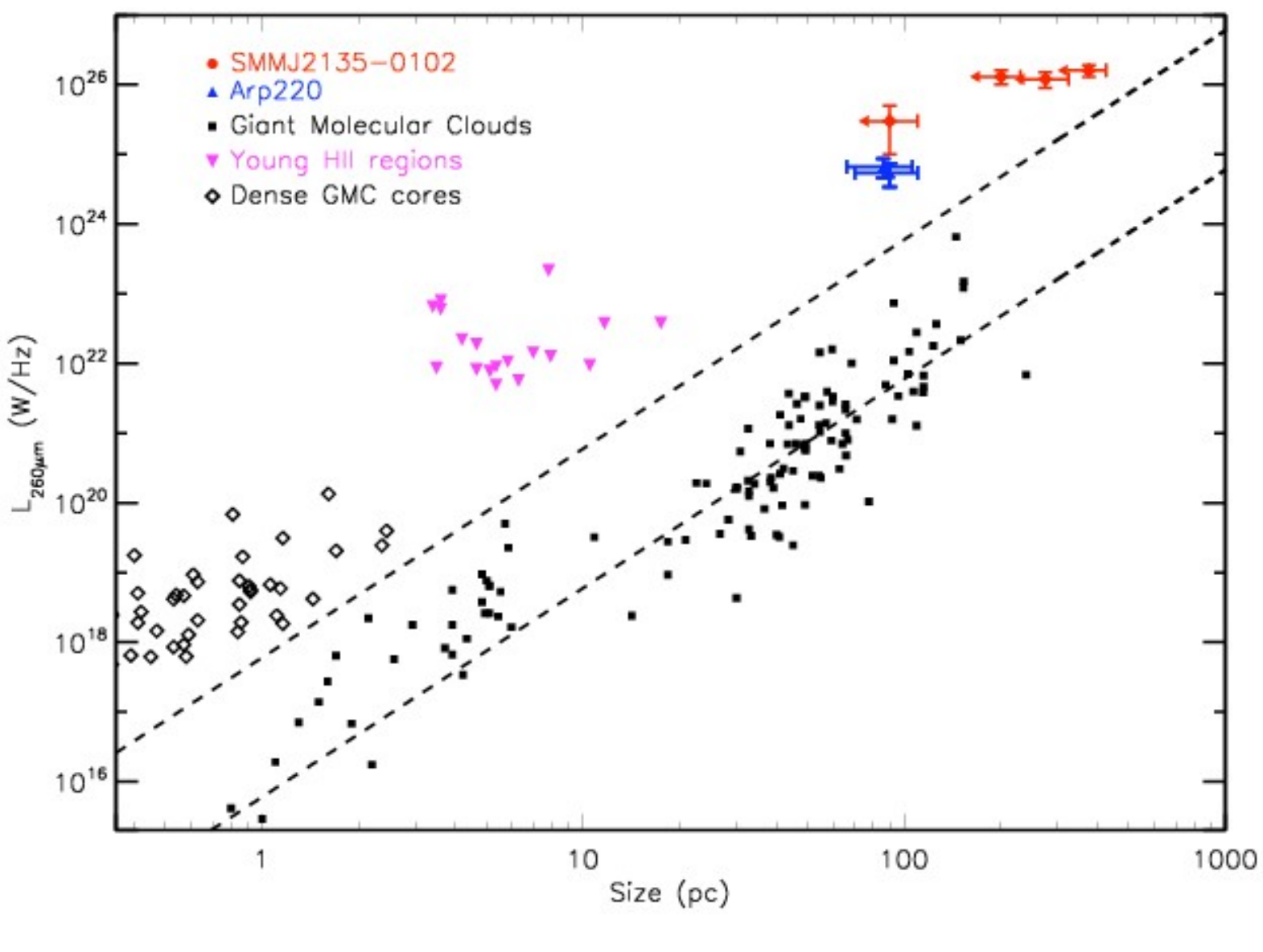


**Figure 4. Relation between size and luminosity of star-forming regions.** Black squares denote the size and 260μm luminosities of giant molecular clouds in the Milky Way and local group[20,22-24], scaled from IRAS. In these comparison samples, we correct the rest-frame 100μm luminosities to 260μm by fitting a modified black-body to the 100μm luminosity at the known temperature (which is derived from $L_{60}/L_{100}$ in each case). We note that, since the peak of a 30K black body spectrum lies at ~170μm, the correction is typically small, with a median $L_{260}/L_{100}$=1.5±0.4. The lower dashed line shows the $L_{260} \propto r^3$ relation for constant energy density, a line with slope=3. The solid red points denote the sizes and luminosities of the star-forming regions in SMMJ2135-0102. Since the luminosities of the star-forming regions within SMMJ2135-0102 are ~100× more luminous at a fixed size compared to local GMCs, the upper dashed line shows the local fit but offset by a factor 100× in luminosity. We also plot the sizes and luminosities of dense cores of galactic GMCs[25] (open diamonds), young (<3Myr) HII regions in Henize 2-10 and M82 (inverted triangles), and the two dominant star-forming regions within the local ULIRG Arp220[28] (blue triangles). In these comparison samples, the conversion between sub-mm flux and rest-frame 260mm luminosity is also calculated using a modified black-body fit at known temperature (Supplementary Information).

## Supplementary Information

### 1. Observations and Data Reduction

*1.1 APEX LABOCA and SABOCA observations*

Observations of the galaxy cluster MACS J2135-0102 ($z_{cl}$=0.325) were made with the LABOCA 870μm bolometer camera[29] on the APEX telescope on 2009 May 8 for a total of 3.2hours (1200 seconds on-source) in excellent conditions (PWV=0.35—0.40mm). We used a 6 arcminute spiral pattern scan, centered at α:21:35:12.706 δ:-01:01:43.27 (J2000). For flux calibration, Mars and Uranus were both observed immediately prior to the science observations. The data were reduced using the minicrush reduction package, which includes temperature drift correction, flat-fielding, opacity correction, bad bolometer masking and de-spiking[29]. The final map appears flat, and has an r.m.s. of 3.5mJy/beam. Including systematic effects, we estimate calibration and fitting uncertainties as ~4% and 6% respectively. Visual inspection of the image shows a bright 30σ source centered at *α*:21:35:11.6 *δ*:-01:02:52.0 with an 870μm flux of 106.0±3.5mJy. Given the uncertainties in the calibration we adopt the flux of the SMG as $S_{870}$=106.5±7.0mJy, and derive $S_{870}$=24.2±7.0mJy for the counter-image.

We followed up SMMJ2135-0102 with the Submillimetre APEX Bolometer CAmera (SABOCA, Siringo et al., in prep.) on the APEX telescope on UT 2009 September 20 and 21. SABOCA is a 37 superconducting Transition Edge Sensing (TES) bolometer array with hexagonal layout and two-beam separation on sky Its filter transmission curve is optimised to cover the 350μm window, and has a central wavelength of 352μm (852 GHz) for flat spectrum sources. SABOCA operates at a temperature of 300mK and is installed in the Cassegrain cabin of APEX. We observed SMMJ2135-0102 in a 20”x20” raster of spirals with 35 seconds duration in order to obtain a fully sampled map. A total of 10 such rasters were used in the final map, corresponding to an on-source integration time of 1400 seconds (the total time, including overheads, was 2.7 hours).

Conditions during the observations were very good, with a stable atmosphere and zenith opacity of $t_{350\mu m}$=0.8 (PWV=0.25mm), determined every hour by skydips. The absolute flux calibration was determined using the primary calibrators Uranus and Neptune, 3 observations each, just before and after SMMJ2135-0102. We used the GILDAS/ASTRO model as a reference to calculate the expected flux, and reduced the data using minicrush. To determine the pointing corrections, we used J2253+161 as a pointing source, but the remaining data still contained some pointing drifts. We therefore reduced each scan individually in Azimuth-Elevation coordinates, which results in a 5 to 7.5σ detection used to register the source. We smoothed the combined map by the 7.5” beam size for presentation purposes, resulting in an effective beam size of 10.6”. We do not find any significant evidence for spatially resolved emission in the unsmoothed map. SMMJ2135-0102 is detected in the beam-smoothed map with $S_{352\mu m}$=530±60mJy, where the uncertainty includes factors from the opacity determination, scatter between the values from the two primary flux calibrators, and Gaussian fitting. We do not include any uncertainties in the planetary flux prediction models of ASTRO, which may add an additional 10 to 15% uncertainty.

*1.2 Optical, Near- and Mid-Infrared Observations*

We used existing *HST* ACS imaging of this cluster[30] to identify a faint $I_{AB}$=23.6±0.2, $K_{AB}$=19.77±0.07 galaxy, extending in a roughly East—West direction, at the position of the sub-mm source. . This cluster has also been observed extensively with the *Spitzer Space Telescope*, and we used the IRAC 3.6,4.5,5.8 & 8.0μm and MIPS 24 & 70μm[31,32] image to identify a mid-infrared counterpart.

Table 1: Optical to far-infrared photometry for SMM J2135-0102 at position *α*:21:35:11.6 *δ*:-01:02:52.0 J2000.

| Band/Filter | Flux | |
|---|---|---|
| $U_{336}$ | <0.1 μJy | HST $U_{336}$ |
| $V_{606}$ | 0.9±0.2 μJy | HST $V_{606}$ |
| $I_{814}$ | 1.4±0.4 μJy | HST $I_{814}$ |
| K | 36±4 μJy | UKIRT K |
| 3.6μm | 0.13±0.02 mJy | SST IRAC Ch1 |
| 4.5μm | 0.21±0.02 mJy | SST IRAC Ch2 |
| 5.8μm | 0.32±0.05 mJy | SST IRAC Ch3 |
| 8.0μm | 0.30±0.05 mJy | SST IRAC Ch4 |
| 24μm | 2.6±0.2 mJy | SST MIPS |

| | | |
|---|---|---|
| 70μm | 6.0±2.6 mJy | SST MIPS |
| 350μm | 530±60 mJy | APEX/SABOCA |
| 434μm | 430±80 mJy | SMA 690GHz |
| 870μm | 106.0±7.0 mJy | APEX/LABOCA |
| 1.2mm | 25.5±4.0 mJy | SMA1.2mm |
| 2.8mm | 1.4±0.3mJy | PdBI 2.8mm |
| 8.6mm | 0.13±0.04 mJy | GBT/Zpec |

Notes: Observed photometry for SMM J2135-0102. To correct for lensing to find the intrinsic fluxes, divide by the amplification factor 32.4±4.5x (Δm=3.8mags). We note that 1mJy corresponds to $m_{AB}$=23.90. We also note that the counter-image of the galaxy is located at $\alpha$:21:35:15.56, $\delta$:-01:03:12.4 (J2000) and is ~3 magnitudes fainter.

*1.3 GBT/Zpectrometer Observations*

We used the Zpectrometer, a wideband spectrometer optimised for CO emission line searches with the Green Bank Telescope's Ka-band receiver, to establish the galaxy's redshift. The Zpectrometer has instantaneous frequency coverage from 25.6 to 36.1GHz with resolution 16 MHz, corresponding to $z$ = 2.2 to 3.5 for the CO J=1-0 line and approximately 150km/s resolution[11] near the band centre. Observations were conducted on 2009 May 19 and 27 in moderate weather. The GBT's subreflector chopped between the receiver's two beams at 0.1Hz. Every 4 minutes we alternated the telescope position between the source and a nearby lensed Lyman Break Galaxy ("The Cosmic Eye" at z=3.07[30]), then differenced the spectra to eliminate residual optical offsets. Each pair was repeated 20 times for a total integration time of 5 hours, approximately equally split between the two sessions. Data reduction was performed with version D of the standard Zpectrometer GBT IDL reduction scripts. No baseline has been removed from the spectrum, and the offset from zero represents the continuum difference between the two sources. The flux and bandpass calibrator was 3C48, and we pointed and focused hourly on 2134-0153. The final spectrum is shown in Fig. 2a. We note that the spectrum is sampled at 75km/s, hence alternate points are independent.

*1.4 PdBI Observations*

We used the six-element IRAM Plateau de Bure Interferometer to observe the redshifted CO(3-2) line and continuum near 103.97GHz. The frequency was tuned to the CO(3–2) rotational transition at $z$ = 2.3259, the systemic redshift of the system as derived from the CO(1-0) (see section 1.3). Observations were made in D configuration in Director's Discretionary Time (DDT) on 2009 May 29 with good atmospheric phase stability (seeing = 0.6"—1.6") and reasonable transparency (PWV = 5–15mm). We observed SMMJ2135-0102 with a total on-source observing time of 4 hrs. The spectral correlator was adjusted to detect the line with a frequency resolution of 2.5MHz across the receivers' 980MHz bandwidth. The overall flux scale for each observing epoch was set on MWC349, with observations of 2134+004 for phase and amplitude calibrations. The data were calibrated, mapped and analyzed using the GILDAS software package. Inspection of the data cube shows an extremely bright detection of CO(3–2) line emission (S/N~300) at the position of SMM J2135-0102 (Figure 2b), confirming the redshift from the CO(1-0) emission. A detailed analysis of the CO kinematics and the full CO ladder will be discussed in a future publication.

*1.5 SMA Observations and Reduction*

345GHz (870μm) observations of the galaxy were carried out with the SMA in its very extended configuration (baselines up to 500m) with 8 antennae, on 2009 July 13 in Directors Discretionary Time (DDT). The galaxy was observed over 9 hours in 10 minutes cuts using quasars 2148+069 and 2225-049 as the gain calibrators. The weather over the first 2 hours was only moderate (and subsequently flagged) but then improved significantly, with the humidity dropping to 10% and $\tau_{225GHz}$~0.1 for the remaining time. These atmospheric conditions resulted in excellent phase stability and reasonable sensitivity ($T_{sys}$ at transit was ~300K). 3C454.3 and Uranus were used as bandpass and flux calibrators, respectively, selecting only the shortest baseline for Uranus, where the emission was not resolved out. The r.m.s. noise in the final map is σ=2.1mJy and the synthesis beam is 0.33"x0.21" at a position angle of 15° East of North. Further 690GHz (434μm) observations were carried out in sub-compact configuration on 2009 September 09 in DDT time. The weather conditions were excellent throughout the 8 hour observation with $\tau_{225GHz}$<0.08. Both Neptune and Callisto, which were within 15 degrees of the target on the observation date, were used as gain calibrators. Observations of nearby bright quasars were interspersed with the source to check the robustness of the gain solution. Neptune, Callisto, Mars and Uranus were used to calibrate both the bandpass and flux. We estimate the absolute flux scale for the 345GHz and 690GHz observations is accurate to 10% and 25%, respectively. The r.m.s. noise in the final 690GHz map is 80mJy and in this configuration, the synthesized beam is 2.9x2.3" at a PA of -15° East of North.

The data were calibrated using the MIR IDL package, adapted for SMA (http://www.cfa.harvard.edu/~cqi/mircook.html). In the 345GHz observations, a more accurate, post-observation baseline solution was applied to improve the delays before doing the standard calibration procedure. The data were then exported to MIRIAD to be imaged and cleaned. A first order polynomial was fit to the channels expected to be line-free (ie. those offset by >1000km/s from the CO(9-8) line in the upper sideband) and subtracted from the visibilities to produce separate line and continuum images.

## 2. Gravitational Lens Modeling

With spectroscopic confirmation that this galaxy is a lensed, multiply-imaged, background galaxy we searched for and identified a counter-image in the multi-wavelength imaging at $\alpha$:21:35:15.56, $\delta$:-01:03:12.4. This source is ~3 magnitudes fainter, as expected, and has almost identical colours as SMMJ2135-0102 (the 850μm flux of the counterpart is $S_{870}$=24.2±7.0mJy). We constructed a gravitational lens model for the galaxy cluster which strongly constrained the total mass in the region responsible for lensing SMMJ2135-0102. A striking multiply imaged blue galaxy at $\alpha$:21:35:11.51 $\delta$:-01:03:33.8 (J2000), is approximately 37" due south of the brightest cluster galaxy (BCG); its multiple images were key for constraining the lens model. We obtained its redshift with three 1.2ks exposures using VLT/FORS in MXU mode on 2006 November 13 as part of program 078.A-0420 in 1" seeing and clear conditions. All three images of the multiply imaged blue galaxy were placed on FORS slits, each yielding continuum with a S/N>5. From the resulting spectrum, a redshift of $z$=2.320+/-0.001 was measured for each of the three images from the features of SiIIλλ1526.7,1533.4; FeIIλ1608, CIVλ1549 and AlIλ1670.8. This triple image, together with the spectroscopically confirmed $z$=2.3259 lensed SMG, were used as primary constraints in the lens modelling for the massive cluster at z=0.325. We note that the multiply imaged blue galaxy is within ~500km/s of the $z$=2.3259 SMG (and 200kpc in projection in the source plane), suggesting that the SMG resides in a small group of high redshift[33]galaxies at.

We used LENSTOOL[34,35] to construct a parametric model of the mass distribution reproducing the triply imaged systems[37] (LENSTOOL optimizes the model by minimizing the location of each image in the source plane). We used a simple model with a single cluster-scale mass component, as well as individual galaxy-scale mass components centered on each cluster member (selected from their *HST V-I* colors). Each component was described by a double Pseudo Isothermal Elliptical (dPIE) mass distribution[36,37] and we assumed that the cluster galaxies follow a scaling relation with constant mass-to-light ratio according to an $L*$ cluster galaxy. We obtained a very good fit using this model, with an r.m.s. of 0.2" between the predicted and observed position of the multiple images. The cluster-scale component is centered 9.7kpc East and 10.2kpc South of the BCG, with an ellipticity of 0.25 and a position angle of -9.7 degrees (East of North). The enclosed mass within an aperture of 250kpc is $M$=3.3±0.3x10$^{14}$$M_{\odot}$ with an Einstein radius of $\theta_e$=34.5±2.0" at $z$=2.32 (Richard et al. 2010 in prep). The best-fit parameters of the dPIE profile are given in Table 3.

**Table 2. Gravitational Lens model parameters**

| | ΔRA (") | ΔDec (") | ε | θ (deg) | $r_{core}$ (kpc) | $r_{cut}$ (kpc) | $v_{disp}$ (km/s) |
|---|---|---|---|---|---|---|---|
| DM halo | 2.1±0.5 | -2.2±0.5 | 0.25±0.02 | -9.7±3 | 99±5 | [1000] | 1294±45 |
| BCG | [0] | [0] | [0.15] | [148] | [0.2] | 152±8 | 268±4 |
| L* galaxies | - | - | - | - | [0.15] | [45] | 192±10 |

Note: Numbers in square brackets are note allowed to vary in the fit. Position angles are clockwise from North. For a complete description on the choice of these parameters, see Richard et al. (2009).

**Table 3. Amplification (μ) and source-plane properties of the star-forming regions within SMMJ2135-0102.**

| Component | $S_{870}$ (mJy) (image plane) | $\mu$ | $S_{870}$ (mJy) (source plane) | *FWHM* (pc) (source plane) |
|---|---|---|---|---|
| A1 | 13.2±2.1 | 9.3±1.2 | 1.42±0.22 | 328±40 x 2050±150 |
| B1 | 14.4±2.1 | 12.9±1.6 | 1.12±0.16 | 229±33 x 1880±150 |
| C1 | 14.5±2.1 | 22.0±2.0 | 0.66±0.10 | 205±24 x 2460±100 |
| D1 | 8.2±2.1 | 31.0±3.9 | 0.30±0.07 | 90±20 x 2000±200 |
| D2 | 6.4±2.1 | 19.9±2.5 | 0.34±0.11 | 98±28 x 2600±160 |
| C2 | 14.9±2.1 | 15.5±2.5 | 0.96±0.13 | 180±28 x 2800±150 |
| B2 | 6.6±2.1 | 5.3±0.7 | 1.25±0.39 | 352±50 x 2700±200 |
| A2 | 8.3±2.1 | 4.7±0.6 | 1.77±0.45 | 452±50 x 2700±200 |

In Fig. 1a we show the true colour *HST* ACS/WFPC2 *VI*-band image of the cluster core, and overlay the contours from the APEX/LABOCA map (contours start at 5σ and are spaced in 5σ intervals up to 30σ). We also overlay the $z$=2.3259 tangential (outer) and radial (inner) critical curves, clearly showing that the brightest sub-mm source is formed from a pair of radial images. Fig. 1b contains a true colour IRAC 3.6, 4.5, 8.0μm image around the lensed galaxy with the 350μm contours from SABOCA in overlay. Using the mapping between the image and source-plane, we derive a total amplification factor of 32.5±4.5x, although this varies from ~5x to ~30x across the galaxy image (Table 2).

**3 Spectral Energy Distribution Analysis**

**3.1 Stellar Mass**

We used the extensive multi-wavelength imaging to estimate the stellar mass by fitting the latest stellar population synthesis models to the observed SED based on the photometry in Table 1[38]. We used a Salpeter 1955 initial mass function (IMF)[39], and considered both a solar and sub-solar ($0.2Z_\odot$) metallicity, and both constant star-formation histories and exponentially decaying ($\tau$) models with *e*-folding decay times of up to 100 Myr. We considered the effects of dust in the modelling by adopting the Calzetti et al. (1994) reddening law[40]. For SMM J2135-0102 the best fit SEDs have a range of ages from 10—30Myr with significant dust extinction, E(B-V)=1.0±0.1, and a stellar mass (corrected for lensing) of $M_{stars}=3\pm2\times10^{10}M_\odot$ (although we caution that there are systematic uncertainties in this estimate due to the unknown star-formation history of the galaxy and the strong dust extinction).

**3.2 Bolometric Luminosity**

We also parameterised the far-infrared SED (including *Spitzer*/MIPS 70μm, SABOCA 350μm, SMA 434μm & 870μm, SMA 1.2mm & PdBI 2.8mm photometry; Table 1) of the galaxy using a modified blackbody spectrum[3] (Fig 3). A single modified black-body fit suggestted a characteristic emission temperature of $T_d$=34±4K, but under-predicted the 70μm flux by a factor >100. To improve the fit, we therefore parameterised the SED using two dust component models, fixing $T_d$=30 and 60K. Integrating the SED we derive a bolometric luminosity (corrected for lensing) of $L_{bol}=1.2\pm0.2\times10^{12}L_\odot$, suggesting a star-formation rate of $SFR=210\pm50M_\odot/yr$[16].

**4 Size and Luminosity of star-forming regions**

To construct the comparison samples in the size and luminosities of local star-forming regions, we exploited a number of IRAS and millimetre studies. To ensure a fair comparison is made between the star-forming regions at $z$~2 and $z$=0, we restricted the comparison samples wherever possible to dust continuum observations where extrapolations to rest-frame 260μm luminosity can be reasonably made (rest-frame 260μm corresponds to observed 870μm at $z$=2.32).

First, we exploited IRAS studies of galactic GMCs. For these following comparison samples, we corrected the rest-frame 100μm luminosities to 260μm by fitting a modified black-body to the 100μm luminosity at the known temperature (which is derived from $L_{60}/L_{100}$ in each case). We note that, since for a temperature of 30K a black-body peaks at ~170μm, the correction is typically small, with a median $L_{260}/L_{100}$=1.5±0.4. First, Scoville et al. (1989)[20] derived far-infrared sizes and luminosities of giant molecular clouds in the first galactic quadrant. In this sample, the clouds have a median diameter of 50±10pc and a median rest-frame 260μm luminosities of $4\times10^{19}$W/Hz. Turning to the outer galaxy, Snell et al. (2002)[22] presented far-infrared sizes and luminosities for molecular clouds with typical sizes 1—4pc and luminosities $1\times10^{17}$W/Hz. We also included two IRAS studies of molecular clouds in the Large Magellanic Cloud (LMC) from Caldwell et al. (2002)[23] and Livanou et al. (2006)[24] who investigated the properties of 73 GMCs with sizes 4—150pc and 260μm luminosities $L_{260mm}=10^{20-24}$W/Hz.

We also exploited millimetre observations of the dense cores of galactic GMCs from Hill et al. (2005)[25] (applying the same procedure as above to correct to rest-frame 260μm luminosity). This sample comprises a survey of 131 star-forming complexes suspected of undergoing massive star-formation. From this parent sample, we extracted 53 star-forming regions where reliable sizes, luminosities and temperatures are available (see Table 4 and 6 of Hill et al. 2005). Finally, we also show the sizes and luminosities of young, dense HII regions in Henize 2—10[41] and M82[43]. In both of these samples, we have extrapolated the VLA 7mm luminosities to rest-frame 260mm using the same technique as above, but we estimate that there is an order of magnitude uncertainty in this correction due to the extrapolation from 7mm and 260μm.

Since the galaxy the we detected is a ULIRG at $z$=2.3259, we also compare the properties of the star-forming regions to the "extreme" starburst observed in the local ULIRGs Arp220, using the sub-compact 690GHz (434μm) SMA observations of the source[42]. In this configuration, the synthesised beam was 1.2"x0.9", corresponding to 470 x 350pc at 79.9Mpc. At these wavelengths, Arp220 comprises two prominent components ($S_{435\mu m}$=1.28±0.38Jy and 0.96±0.29Jy for the western and eastern structures respectively), both of which are unresolved in the SMA map. We corrected the corresponding luminosities to rest-frame 260μm using their predicted temperatures ($T_{d,west}$~100K and $T_{d,east}$~60K) and derived $L_{260}/L_{435}$=4.6 and $L_{260}/L_{435}$=7.2 for the western and eastern components respectively, resulting in $L_{260,west}$=4.5±2.0x$10^{24}$ W/Hz and $L_{260,east}$=5.5±2.0x$10^{24}$ W/Hz. Although these components are unresolved at 434μm, higher resolution (0.2") SMA observations at 870μm have resolved the structures on ~50pc scales[28]. In Fig. 4 we therefore show the sizes as resolved at 870μm, but caution that this assumes that there is no significant size difference between 434μm and 850μm.

**Additional References**